\documentclass[final,5p,twocolumn,number]{elsarticle}
\usepackage[T1]{fontenc}
\usepackage{lmodern}
\usepackage{microtype}
\usepackage{amsmath,amsfonts}
\usepackage{graphicx}
\usepackage[hidelinks]{hyperref}
\journal{Physics Letters A}
\begin{document}
\begin{frontmatter}
\title{A simple solvable \texorpdfstring{$Z_N$}{Z\string_N} Hamiltonian in any dimension}
\author[anu]{Yuguan Li}
\ead{yuguan.li@anu.edu.au}
\author[anu]{Murray T. Batchelor\corref{cor}}
\ead{murray.batchelor@anu.edu.au}
\cortext[cor]{Corresponding author.}
\affiliation[anu]{organization={Mathematical Sciences Institute, Australian National University},
  citysep={}, statesep={},
  city={Canberra}, state={ACT}, postcode={2601}, country={Australia}}
\begin{abstract}
We construct local $Z_N$ clock Hamiltonians from Baxter-Fendley chains coupled according to an arbitrary finite oriented simple transverse graph. Bounded edge lengths and vertex degrees give finite-range multispin interactions, with $2d$-site bulk terms on a $d$-dimensional rectangular lattice.
For open longitudinal boundaries, an explicit nonlocal unitary transformation decouples the chains and gives a free parafermion spectrum. On rectangular lattices this holds for any combination of open and periodic transverse boundaries. For periodic longitudinal boundaries, neighbouring chain charges fix the closing-bond twists. In each joint charge sector, the energies are sums of the corresponding twisted periodic-chain energies. As an example, we construct a three-dimensional array of chains perpendicular to a Penrose rhombus tiling.
\end{abstract}
\end{frontmatter}
\section{Introduction}
In this article, we construct a class of local $Z_N$ clock Hamiltonians in
arbitrary dimensions with free parafermion spectra for open longitudinal
boundaries. Each bond interaction acts on two adjacent sites of one chain
and one site on each neighbouring chain.
The transverse couplings between chains can follow the edges of any
finite oriented simple graph. This freedom extends the construction to
quasicrystals, as we illustrate with an array of chains perpendicular
to a Penrose rhombus tiling.

In 1989 Baxter obtained the spectrum of a one-dimensional $Z_N$ clock
Hamiltonian with fixed boundary spins and raised the question of its
spectrum with periodic boundaries~\cite{Baxter1989}. Fendley subsequently 
constructed shift operators connecting eigenstates of the open chain and
derived its free parafermion spectrum algebraically~\cite{Fendley2014}.
We refer to this model as the Baxter-Fendley $Z_N$ clock chain.
A review of these developments is given in Ref.~\cite{BatchelorHenryLu2023}.
A study of the periodic chain 
for $N>2$ found numerical evidence that closing the chain can change
the bulk ground-state energy per site~\cite{AlcarazBatchelor2018}.
For the open chain, exceptional points have been located 
where quasienergies coalesce as the field is varied in the complex
plane~\cite{HenryBatchelor2023}. Unsupervised
learning has been used to detect phase transitions from the quasienergies~\cite{LiuSunBatchelor2026}.
Recently an exact finite-size solution for the periodic chain in terms of
polynomial spectral equations in each $Z_N$ charge
sector has been obtained~\cite{LiLiuBatchelor2026}. 
For $N>2$, the unrestricted open-chain free-parafermion occupation rule generally fails for periodic boundaries. 
At $N=2$, the periodic chain remains solvable by free fermions, with the appropriate parity-dependent boundary conditions.

Multispin chains with free fermionic and parafermionic spectra have been
constructed using exchange-algebra methods~\cite{AP2020Rapid,AP2020Algebra}.
Related work establishes spectral correspondences with $XY$ chains and
inhomogeneous Ising chains~\cite{AP2021XY,APS2023Ising}, and extends the
exchange algebra to nonuniform interaction ranges~\cite{Alcaraz2026Ranges}.
Free-parafermion spectra have also been obtained from a quantum Frobenius relation for the quantized six-vertex model at odd roots of unity~\cite{Inoue2026}. Kuniba and Oikawa subsequently realized arbitrary finite indifference frustration graphs within this framework~\cite{Kuniba2026}.
Here the higher-dimensional construction retains the nearest-neighbour
Baxter-Fendley exchange algebra while changing its local realization on
the physical lattice.
We give two- and three-dimensional Hamiltonians and then
present the general construction in Section~\ref{sec:general}. With open
longitudinal boundaries, a unitary transformation reduces the
Hamiltonian to independent open Baxter-Fendley chains for any choice of open or
periodic transverse boundaries~\cite{Fendley2014}. With periodic longitudinal
boundaries, the charges of neighbouring chains determine the twist
on each chain. In each joint charge sector, the energies are sums of the
corresponding one-dimensional periodic-chain energies~\cite{LiLiuBatchelor2026}.
Section~\ref{sec:penrose} gives the multispin interactions at the
different vertices of the Penrose tiling~\cite{Penrose1979}.
Earlier work on this tiling includes
Korepin's application of Baxter's planar eight-vertex
model~\cite{Baxter1978,Korepin1987} and the classical dimer model~\cite{FlickerSimonParameswaran2020}.
The solvability mechanism considered here is different. An explicit unitary transformation reduces the quantum Hamiltonian to Baxter-Fendley chains, without requiring the transverse graph to arise from a rapidity-line construction.

\section{The two-dimensional model}
\begin{figure}[htbp]
\centering
\includegraphics[width=0.78\columnwidth]{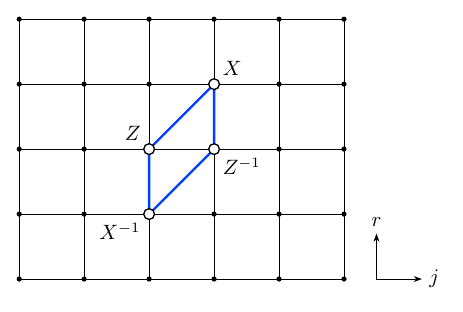}
\caption{Two-dimensional lattice with open boundaries. The four open
circles mark the sites of one interaction in Eq.~\eqref{eq:hamiltonian-2d}.
The blue lines outline the parallelogram formed by these sites.
(Illustrative diagram: OpenAI Codex (GPT-6) assisted with the TikZ 
code for the outline, which was reviewed and verified by the authors.)
}
\label{fig:open-lattice}
\end{figure}

Consider $M$ rows of $L$ sites, with $r$ labelling the row and $j$ the
position along it. The Hamiltonian is
\begin{equation}
\begin{aligned}
H_{\mathrm{2D}}={}&\sum_{r,j}h_{r,j}X_{r,j}\\
 &+\sum_{r,j}J_{r,j}\,X_{r+1,j+1}X_{r-1,j}^{-1}
                      Z_{r,j}Z_{r,j+1}^{-1}.
\end{aligned}
\label{eq:hamiltonian-2d}
\end{equation}
Here $N,L\ge 2$, $1\le r\le M$, and $1\le j\le L$.
Each coupling $h_{r,j}$ or $J_{r,j}$ can be chosen independently as an
arbitrary complex number.
The Hamiltonian is generally non-Hermitian, although the decoupling transformation introduced below is unitary.
Each bulk interaction couples a longitudinal bond in row $r$ to one
site in each neighbouring row, as shown in Fig.~\ref{fig:open-lattice}.
In all dimensions indices wrap around in a periodic direction. Each
periodic transverse direction has at least three sites. For open longitudinal
boundaries, omit the $j=L$ interaction (equivalently, set $J_{r,L}=0$).
At an open transverse
boundary, omit the $X$ factor belonging to the missing row.

Each site carries an $N$-state clock. In the local basis
$\{|s\rangle:s=0,\ldots,N-1\}$, the clock operator $Z$ and the shift
operator $X$ act as
\begin{equation}
Z|s\rangle=\omega^s|s\rangle,\qquad X|s\rangle=|s+1\bmod N\rangle,
\end{equation}
where $\omega=\exp(2\pi i/N)$. They are unitary and obey the Weyl
relations
\begin{equation}
ZX=\omega XZ,\qquad X^N=Z^N=\mathbf{1}.
\end{equation}
Thus $X^{-1}=X^\dagger$ and $Z^{-1}=Z^\dagger$. Operators on distinct
sites commute. For the one-dimensional $Z_N$ clock chain, 
the Hamiltonian can be expressed in terms of $N-1$ types of generators
of a coupled Temperley-Lieb algebra~\cite{AddertonBatchelor2026}.

\subsection{Open boundary conditions}
Take open boundaries in both directions. In the joint eigenbasis of
the shift operators, write
$X_{r,j}|\boldsymbol n\rangle_X=\omega^{n_{r,j}}|\boldsymbol n\rangle_X$,
with $n_{r,j}\in\mathbb Z_N$. Define the diagonal unitary operator $U$ by
\begin{equation}
\begin{aligned}
U|\boldsymbol n\rangle_X&=\omega^{F(\boldsymbol n)}|\boldsymbol n\rangle_X,\\
F(\boldsymbol n)&=\sum_{r=1}^{M-1}\sum_{\ell=1}^{L}
                  n_{r,\ell}\sum_{k=1}^{\ell} n_{r+1,k}.
\end{aligned}
\label{eq:open-unitary}
\end{equation}
Since $U$ commutes with every $X_{r,j}$, it suffices to compute its
action on a longitudinal bond $B_{r,j}=Z_{r,j}Z_{r,j+1}^{-1}$, with
$1\le j<L$. The Weyl relation implies that $B_{r,j}$ lowers $n_{r,j}$
by one and raises $n_{r,j+1}$ by one. The $X$-basis labels $n_{r,j}$,
charge labels, and changes in $F$ are evaluated modulo $N$. In
Eq.~\eqref{eq:open-unitary}, the partial sums over row $r+1$ then
differ by $n_{r+1,j+1}$. In the term coupling rows $r-1$ and $r$,
only the partial sum ending at $j$ changes, contributing $-n_{r-1,j}$
to the change in $F$. Writing this change as $\Delta F$, conjugation
in the order $U^\dagger B_{r,j}U$ gives the phase $\omega^{-\Delta F}$.
Thus
\begin{equation}
U^\dagger B_{r,j}U
 =X_{r+1,j+1}^{-1}X_{r-1,j}B_{r,j},
\label{eq:open-bond}
\end{equation}
with the factor from a missing row omitted at $r=1$ or $r=M$.
These factors cancel the transverse $X$ factors in
Eq.~\eqref{eq:hamiltonian-2d}, giving
\begin{equation}
\begin{aligned}
U^\dagger H_{\mathrm{2D}}U &= \sum_{r=1}^{M} H_r,\\
H_r &= \sum_{j=1}^{L} h_{r,j}X_{r,j}
     + \sum_{j=1}^{L-1} J_{r,j}B_{r,j}.
\end{aligned}
\label{eq:open-chains}
\end{equation}
Each $H_r$ is the Hamiltonian of the open Baxter-Fendley chain on row $r$.
The row Hamiltonians therefore act on independent tensor factors, and
the energies of the array are sums of independently chosen open-chain
energies~\cite{Baxter1989,Fendley2014}.
Each open-chain energy is a sum of $L$ quasienergies, each multiplied independently by an $N$th root of unity.

\subsection{Periodic boundary conditions}
Close each row by identifying $j=L+1$ with $j=1$, and write
$B_{r,L}=Z_{r,L}Z_{r,1}^{-1}$. If the transverse direction is also
periodic, extend the row sum in Eq.~\eqref{eq:open-unitary} to $r=M$,
with row $M+1$ identified with row $1$. For open transverse boundaries
use Eq.~\eqref{eq:open-unitary} as written. In either case
Eq.~\eqref{eq:open-bond} applies to all bonds with $j<L$.

The closing bond transfers one unit from $n_{r,L}$ to $n_{r,1}$.
The charge label of row $r$ is $q_r=\sum_{j=1}^L n_{r,j}\bmod N$. The partial sums in
Eq.~\eqref{eq:open-unitary} now give $n_{r+1,1}-q_{r+1}$ from
row $r+1$ and $q_{r-1}-n_{r-1,L}$ from row $r-1$.
Consequently, after the transverse $X$ factors cancel,
\begin{equation}
U^\dagger\!\left(X_{r+1,1}X_{r-1,L}^{-1}B_{r,L}\right)U
= Q_{r+1}Q_{r-1}^{-1}B_{r,L},
\label{eq:periodic-bond}
\end{equation}
where $Q_r=\prod_{j=1}^L X_{r,j}$ is the $Z_N$ charge operator
of row $r$~\cite{Fendley2014,LiLiuBatchelor2026}.
Each longitudinal bond preserves the sum of the two $X$-basis labels at its ends, while the transverse $X$ factors leave all such labels unchanged. Hence every $Q_r$ commutes with the Hamiltonian.

In the joint charge sector $\mathcal H_{\boldsymbol q}$ where every $Q_r$
has eigenvalue $\omega^{q_r}$, the transformed Hamiltonian is
\begin{equation}
\begin{aligned}
\left.U^\dagger H_{\mathrm{2D}}U\right|_{\mathcal H_{\boldsymbol q}}
 &=\sum_{r=1}^M \left.H_r^{\mathrm{per}}(\theta_r)\right|_{q_r},\\
H_r^{\mathrm{per}}(\theta_r)&=H_r+\theta_r J_{r,L}B_{r,L},\\
\theta_r&=\omega^{q_{r+1}-q_{r-1}},
\end{aligned}
\label{eq:periodic-chains}
\end{equation}
with $H_r$ as in Eq.~\eqref{eq:open-chains}. For periodic transverse
boundaries the row labels are taken modulo $M$. For open transverse
boundaries set $q_0=q_{M+1}=0$ and omit the corresponding $X$ factors
in Eq.~\eqref{eq:periodic-bond}.

Let $\epsilon_{r,\nu_r}(q_r,\theta_r)$ be a many-body energy eigenvalue of the
periodic chain on row $r$ in charge sector $q_r$ with closing-bond coupling
$\theta_r J_{r,L}$. To use the bond convention of
Ref.~\cite{LiLiuBatchelor2026}, reverse the site order. The on-site coupling
at $k$ is then $h_{r,L+1-k}$, the bond at $k<L$ has coupling
$J_{r,L-k}$, and the closing-bond coupling is $\theta_r J_{r,L}$.
Site reversal preserves $q_r$.
The full array spectrum is
\begin{equation}
E_{\boldsymbol q,\boldsymbol\nu}
=\sum_{r=1}^M \epsilon_{r,\nu_r}(q_r,\theta_r).
\label{eq:periodic-spectrum}
\end{equation}
For each choice of $\boldsymbol q$, the neighbouring charges fix the
twists and the eigenvalue labels $\nu_r$ can then be chosen independently.
Taking all $\boldsymbol q$ gives the full spectrum with algebraic multiplicities.
Closing only the transverse direction changes $U$ but leaves
Eq.~\eqref{eq:open-chains} and its full free parafermion spectrum intact.
For $N>2$ the longitudinal closing bonds generally invalidate the
independent open-chain occupation rule. 
Eq.~\eqref{eq:periodic-spectrum} applies with either open or periodic transverse boundaries.

\section{The three-dimensional model}
\begin{figure}[htbp]
\centering
\includegraphics[width=0.93\columnwidth]{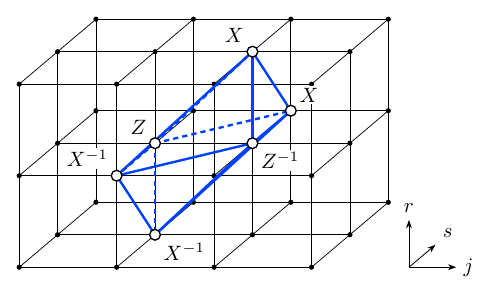}
\caption{Three-dimensional lattice with open boundaries. The six open
circles mark the sites of one interaction in Eq.~\eqref{eq:hamiltonian-3d}.
The blue lines outline the octahedron formed by these sites, with
hidden edges dashed. 
(Illustrative diagram: OpenAI Codex (GPT-6) assisted
with the TikZ code for the outline and hidden edges, which was reviewed
and verified by the authors.)
}
\label{fig:open-lattice-3d}
\end{figure}

On an $R\times S\times L$ lattice, let $(r,s)$ label a chain and $j$
the position along it. With the same clock and shift operators as in
two dimensions, the Hamiltonian is
\begin{equation}
\begin{aligned}
H_{\mathrm{3D}}={}&\sum_{r,s,j}h_{r,s,j}X_{r,s,j}\\
 &+\sum_{r,s,j} J_{r,s,j} C_{r,s,j} Z_{r,s,j}Z_{r,s,j+1}^{-1},\\
C_{r,s,j}={}&X_{r+1,s,j+1}X_{r-1,s,j}^{-1}\\
 &\times X_{r,s+1,j+1}X_{r,s-1,j}^{-1}.
\end{aligned}
\label{eq:hamiltonian-3d}
\end{equation}
The sums run over $1\le r\le R$, $1\le s\le S$, and $1\le j\le L$.
The couplings $h_{r,s,j}$ and $J_{r,s,j}$ are arbitrary complex numbers.
Each bulk interaction acts on a longitudinal bond and one site on
each of the four neighbouring chains, as shown in Fig.~\ref{fig:open-lattice-3d}.
For open longitudinal boundaries, set $J_{r,s,L}=0$. At an open transverse
boundary, omit the $X$ factors on missing chains. In a periodic direction,
take the corresponding index modulo its length.

\subsection{Open boundary conditions}
Take open boundaries in all three directions. Write $a=(r,s)$ for a
chain and $X_{a,j}|\boldsymbol n\rangle_X=\omega^{n_{a,j}}
|\boldsymbol n\rangle_X$. For each pair of adjacent chains, write
$a\to b$ if $b=(r+1,s)$ or $b=(r,s+1)$. The sum below includes only
pairs inside the lattice. Define the diagonal unitary operator $U$ by
\begin{equation}
\begin{aligned}
U|\boldsymbol n\rangle_X&=\omega^{F(\boldsymbol n)}|\boldsymbol n\rangle_X,\\
F(\boldsymbol n)&=\sum_{a\to b}\sum_{\ell=1}^{L}
 n_{a,\ell}\sum_{k=1}^{\ell}n_{b,k}.
\end{aligned}
\label{eq:3d-open-unitary}
\end{equation}
Each transverse direction thus contributes a sum of the same form as
in Eq.~\eqref{eq:open-unitary}.

For $j<L$, the bond $B_{a,j}=Z_{a,j}Z_{a,j+1}^{-1}$ lowers $n_{a,j}$
by one and raises $n_{a,j+1}$ by one. For each pair $a\to b$, the
partial sums ending at $j+1$ and $j$ differ by $n_{b,j+1}$.
For each pair $c\to a$, only the partial sum ending at $j$ changes,
contributing $-n_{c,j}$ to the change in $F$. Thus
\begin{equation}
\begin{aligned}
\Delta F={}&n_{r+1,s,j+1}-n_{r-1,s,j}\\
           &+n_{r,s+1,j+1}-n_{r,s-1,j}.
\end{aligned}
\end{equation}
Terms from missing neighbours are omitted. The phase $\omega^{-\Delta F}$
in $U^\dagger B_{a,j}U$ cancels the transverse factors in
Eq.~\eqref{eq:hamiltonian-3d}, giving
\begin{equation}
 U^\dagger B_{a,j}U=C_{a,j}^{-1}B_{a,j},\qquad
 U^\dagger(C_{a,j}B_{a,j})U=B_{a,j}.
\end{equation}
Here we used the fact that $U$ commutes with every $X_{a,j}$, which
also leaves the on-site terms invariant. The transformed Hamiltonian is
\begin{equation}
\begin{aligned}
U^\dagger H_{\mathrm{3D}}U &=\sum_{r=1}^{R}\sum_{s=1}^{S}H_{r,s},\\
H_{r,s}&=\sum_{j=1}^{L}h_{r,s,j}X_{r,s,j}
       +\sum_{j=1}^{L-1}J_{r,s,j}B_{(r,s),j}.
\end{aligned}
\label{eq:3d-open-chains}
\end{equation}
Each $H_{r,s}$ acts only on chain $(r,s)$. The array is therefore
unitarily equivalent to $RS$ independent open Baxter-Fendley chains. Each array
energy is a sum of open-chain energies~\cite{Baxter1989,Fendley2014}.

\subsection{Periodic boundary conditions}
Close each longitudinal chain $a=(r,s)$ with the bond
$B_{a,L}=Z_{a,L}Z_{a,1}^{-1}$. In each periodic transverse direction,
include the pairs across that boundary in Eq.~\eqref{eq:3d-open-unitary}
and take the corresponding index modulo its length. For an open
transverse boundary, omit pairs involving a missing chain.

The $Z_N$ charge operator of chain $a$ is $Q_a=\prod_{j=1}^{L}X_{a,j}$, with eigenvalue
$\omega^{q_a}$, where $q_a=\sum_j n_{a,j}\bmod N$~\cite{Fendley2014,LiLiuBatchelor2026}.
Every bond transfers one unit within a single chain, so all $Q_a$ are
conserved. The closing bond transfers one unit from site $L$ to site $1$.
In the exponent of Eq.~\eqref{eq:3d-open-unitary}, each pair $a\to b$
then contributes $n_{b,1}-q_b$ to $\Delta F$, while each pair $c\to a$
contributes $q_c-n_{c,L}$. The site-dependent factors in
$\omega^{-\Delta F}$ cancel $C_{a,L}$, giving
\begin{equation}
\begin{aligned}
U^\dagger(C_{a,L} B_{a,L})U &= \Theta_{r,s} B_{a,L},\\
\Theta_{r,s} &= Q_{r+1,s}Q_{r-1,s}^{-1}
                 Q_{r,s+1}Q_{r,s-1}^{-1}.
\end{aligned}
\label{eq:3d-periodic-bond}
\end{equation}
At an open transverse boundary, omit the charge operators of missing neighbours.

Fix the charge labels of all $RS$ chains. In this joint charge sector $\Theta_{r,s}$ is
the scalar twist
\begin{equation}
\theta_{r,s}=\omega^{q_{r+1,s}-q_{r-1,s}+q_{r,s+1}-q_{r,s-1}},
\label{eq:3d-twist}
\end{equation}
with missing neighbours omitted. Since the interior bonds transform as
in the open chain, the Hamiltonian in this sector becomes
\begin{equation}
\left.U^\dagger H_{\mathrm{3D}}U\right|_{\mathcal H_{\boldsymbol q}}
=\sum_{r,s}\left.\bigl(H_{r,s}
+\theta_{r,s}J_{r,s,L}B_{(r,s),L}\bigr)\right|_{q_{r,s}},
\label{eq:3d-periodic-chains}
\end{equation}
where $H_{r,s}$ is the open-chain Hamiltonian in Eq.~\eqref{eq:3d-open-chains}.
For each chain, take an eigenvalue $\epsilon_{r,s,\nu_{r,s}}$ from the
one-dimensional periodic solution~\cite{LiLiuBatchelor2026}, using its
own charge sector $q_{r,s}$ and closing-bond coupling $\theta_{r,s}J_{r,s,L}$.
The three-dimensional energies are
\begin{equation}
E_{\boldsymbol q,\boldsymbol\nu}
=\sum_{r=1}^{R}\sum_{s=1}^{S}
 \epsilon_{r,s,\nu_{r,s}}(q_{r,s},\theta_{r,s}).
\label{eq:3d-periodic-spectrum}
\end{equation}
Once $\boldsymbol q$ fixes all twists, the eigenvalue labels $\nu_{r,s}$
can be chosen independently. Taking these sums over all joint charge
sectors gives the full spectrum, including algebraic multiplicities.

\section{Arbitrary transverse graphs}
\label{sec:general}
Each longitudinal bond may carry any number of transverse $X$ factors.
Let the vertices of a finite simple graph label chains of length $L$ and fix
one orientation of each edge for all $j$. The Hamiltonian is
\begin{equation}
\begin{aligned}
H={}&\sum_{a,j}h_{a,j}X_{a,j}
 +\sum_{a,j}J_{a,j}C_{a,j}Z_{a,j}Z_{a,j+1}^{-1},\\
C_{a,j}={}&\prod_{b:\,a\to b}X_{b,j+1}
             \prod_{c:\,c\to a}X_{c,j}^{-1}.
\end{aligned}
\label{eq:d-hamiltonian}
\end{equation}
Each edge $a\to b$ contributes $X_{b,j+1}$ on the bond of chain $a$
and $X_{a,j}^{-1}$ on that of chain $b$.
On a rectangular lattice, orient edges along the $d-1$ positive
transverse axes to recover the $2d$-site bulk interaction. Open boundaries
omit edges across them. A single chain has $C_{a,j}=1$. Bounded edge
lengths and vertex degrees give a uniform bound on the interaction range
and on the number of sites in each term.

Extend the unitary of Eq.~\eqref{eq:3d-open-unitary} to the transverse
graph by summing over all its directed edges. In the joint eigenbasis of
the shift operators it acts as
\begin{equation}
\begin{aligned}
 U|\boldsymbol n\rangle_X&=\omega^{F(\boldsymbol n)}|\boldsymbol n\rangle_X,\\
 F(\boldsymbol n)&=\sum_{a\to b}\sum_{\ell=1}^{L}
                     n_{a,\ell}\sum_{k=1}^{\ell}n_{b,k}.
\end{aligned}
\label{eq:d-unitary}
\end{equation}
No acyclicity condition is imposed on the transverse orientation.
Writing $B_{a,j}=Z_{a,j}Z_{a,j+1}^{-1}$ for an interior bond, we obtain
\begin{equation}
\begin{aligned}
 U^\dagger X_{a,j}U &= X_{a,j} &&(1\le j\le L),\\
 U^\dagger B_{a,j}U &= C_{a,j}^{-1}B_{a,j} &&(1\le j<L),\\
 U^\dagger(C_{a,j}B_{a,j})U &= B_{a,j} &&(1\le j<L).
\end{aligned}
\label{eq:d-interior-bond}
\end{equation}
With open longitudinal boundaries ($J_{a,L}=0$) we obtain
\begin{equation}
\begin{aligned}
 U^\dagger H U &= \sum_a H_a^{\mathrm{open}},\\
 H_a^{\mathrm{open}} &= \sum_{j=1}^{L}h_{a,j}X_{a,j}
                    +\sum_{j=1}^{L-1}J_{a,j}B_{a,j}.
\end{aligned}
\label{eq:d-open-chains}
\end{equation}
Each term $H_a^{\mathrm{open}}$ acts on a single Baxter-Fendley chain, so the
energies are
\begin{equation}
E_{\boldsymbol m}=\sum_a\sum_{k=1}^{L}
 \omega^{m_{a,k}}\varepsilon_{a,k}.
\label{eq:d-open-spectrum}
\end{equation}
Here $\varepsilon_{a,k}$ are the quasienergies of the one-dimensional
open Baxter-Fendley chain with couplings $h_{a,j}$ and $J_{a,j}$, and each
$m_{a,k}$ ranges independently from $0$ to $N-1$ on the full Hilbert
space~\cite{Fendley2014,AlcarazBatchelorLiu2017}.
The index $k$ labels a single mode on chain $a$.
Equation~\eqref{eq:d-open-spectrum} lists the eigenvalues with algebraic multiplicity, including at parameter values where the Hamiltonian is not diagonalizable. The unitary equivalence in Eq.~\eqref{eq:d-open-chains} remains valid at such points.

The open-boundary construction also has a direct interpretation in the
frustration-graph framework~\cite{Mann2025}.  Write
\begin{equation}
 A_{a,j}=X_{a,j},\qquad D_{a,j}=C_{a,j}B_{a,j},
 \label{eq:graph_generators}
\end{equation}
where $1\leq j\leq L$ for $A_{a,j}$ and $1\leq j<L$ for
$D_{a,j}$.  Equation~(21) maps these operators term by term to the
field and bond operators of independent Baxter-Fendley chains.  Their exchange
relations are therefore
\begin{equation}
 \begin{aligned}
 [A_{a,j},A_{b,k}]&=[D_{a,j},D_{b,k}]=0,\\
 A_{a,j}D_{b,k}
 &=\omega^{\delta_{ab}(\delta_{j,k+1}-\delta_{j,k})}
 D_{b,k}A_{a,j}.
 \end{aligned}
 \label{eq:graph_exchange}
\end{equation}
These relations do not depend on the transverse graph.
Up to reversing the generator ordering, the algebra of each open chain is
the $p=1$ exchange algebra discussed in Ref.~\cite{AP2020Algebra}.
The larger physical support of the dressed interactions does not enlarge
their frustration graph.

The vertices of the frustration graph represent Hamiltonian terms rather
than physical sites.  For $N>2$, orient an edge from $P$ to $Q$ when
$PQ=\omega QP$.  For nonzero couplings, each open chain contributes the
directed path
\begin{equation}
 A_{a,1}\leftarrow D_{a,1}\leftarrow A_{a,2}
 \leftarrow\cdots\leftarrow D_{a,L-1}\leftarrow A_{a,L}.
 \label{eq:graph_path}
\end{equation}
Consequently,
\begin{equation}
 G_{\mathrm{frust}}=\bigsqcup_a\overrightarrow{P}_{2L-1},
 \label{eq:graph_disjoint_paths}
\end{equation}
where $\overrightarrow{P}_{2L-1}$ denotes a consistently oriented path
on $2L-1$ vertices.  At $N=2$ the arrows carry no additional information,
and the underlying undirected paths describe the anticommutation
relations.  Vanishing couplings delete the corresponding vertices.

These paths are oriented indifference graphs, the class for which
Ref.~\cite{Mann2025} gives a free-parafermion construction.  
For odd $N$, these frustration graphs also admit the Q6V realization of Kuniba and Oikawa~\cite{Kuniba2026}.
Thus the transverse graph, including a Penrose graph, specifies the physical
support of the interactions but does not change the open-chain
frustration graph.  The present construction supplies an explicit
unitary realization of this chain algebra by local multispin
interactions on transverse graphs with bounded edge lengths and degrees.

Despite this locality, the decoupling transformation is nonlocal along the chains. 
In particular,
\begin{equation}
U^\dagger Z_{a,j}U
=
\left(
\prod_{b:a\to b}\prod_{k=1}^{j}X_{b,k}
\right)
\left(
\prod_{c:c\to a}\prod_{k=j}^{L}X_{c,k}
\right)
Z_{a,j}.    
\end{equation}
A local clock operator is therefore transformed into strings on neighbouring chains. The reduction to independent chains determines the spectrum, but original-spin correlation functions generally require string correlation functions in the chain representation.

For periodic longitudinal boundaries ($j+L\equiv j$), write
$B_{a,L}=Z_{a,L}Z_{a,1}^{-1}$. The closing interaction transforms as
\begin{equation}
\begin{aligned}
 U^\dagger(C_{a,L}B_{a,L})U &= \Theta_a B_{a,L},\\
 \Theta_a &= \prod_{b:\,a\to b}Q_b\prod_{c:\,c\to a}Q_c^{-1},\\
 Q_a &= \prod_{j=1}^L X_{a,j}.
\end{aligned}
\label{eq:d-periodic-bond}
\end{equation}
In the joint charge sector $\mathcal H_{\boldsymbol q}$ where each $Q_a$ has
eigenvalue $\omega^{q_a}$, the twist on chain $a$ is
\begin{equation}
\theta_a=\omega^{\sum_{b:\,a\to b}q_b-\sum_{c:\,c\to a}q_c}.
\label{eq:d-twist}
\end{equation}

Charge-dependent boundary twists also arise when nonlocal transformations
decouple gauge-coupled fermion chains~\cite{SchulzShastry1998,SchulzShastry1999}.
Here the twists are fixed by the neighbouring $Z_N$ chain charges according
to the oriented transverse graph.
The chains decouple as
\begin{equation}
\begin{aligned}
 \left.U^\dagger H U\right|_{\mathcal H_{\boldsymbol q}}
  &=\sum_a\left.H_a^{\mathrm{per}}(\theta_a)\right|_{q_a},\\
 H_a^{\mathrm{per}}(\theta_a)
  &=H_a^{\mathrm{open}}+\theta_a J_{a,L}B_{a,L}.
\end{aligned}
\label{eq:d-periodic-chains}
\end{equation}
Choosing one energy from each chain in its own sector gives
\begin{equation}
E_{\boldsymbol q,\boldsymbol\nu}
 =\sum_a\epsilon_{a,\nu_a}(q_a,\theta_a).
\label{eq:d-periodic-spectrum}
\end{equation}
Each $\epsilon_{a,\nu_a}$ is an energy eigenvalue of the periodic chain
in charge sector $q_a$ with closing-bond coupling $\theta_a J_{a,L}$.
The periodic spectral equations
of Ref.~\cite{LiLiuBatchelor2026} apply for nonzero on-site couplings.
The sector characteristic polynomials extend to zero couplings by
continuity. For $A$ chains, each of the $N^A$ joint charge sectors gives
$(N^{L-1})^A$ energies, since $\nu_a=1,\ldots,N^{L-1}$ independently.
This accounts for all $N^{AL}$ energies with algebraic multiplicity,
even when the Hamiltonian is defective.

\subsection{The Ising case}

For $N=2$, the operators $X$ and $Z$ are Pauli matrices and the
Hamiltonian is Hermitian when all $h_{a,j}$ and $J_{a,j}$ are real.
The unitary transformation then gives independent transverse-field
Ising chains for open longitudinal boundaries. The unitary also satisfies
$U=U^\dagger$ and $U^2=I$. It is a Clifford transformation, since its
action maps the Pauli generators to Pauli strings, although it remains
nonlocal along the chains.
Clifford transformations that produce local integrable Hamiltonians,
including transformations that send some individual operators to nonlocal
strings, were studied by Jones and Linden~\cite{JonesLinden2022}.
A different mechanism for embedding one-dimensional fermionic problems
in higher-dimensional spin-$1/2$ models was given by Pujari~\cite{Pujari2025}.
The present construction supplies an explicit $Z_N$ unitary reduction on
arbitrary transverse graphs.

For the regular two-dimensional construction in Eq.~\eqref{eq:hamiltonian-2d}, 
the bulk interaction has a direct relation to the Wen-plaquette model~\cite{Wen2003,YuKouWen2008}.
Relabel the sites by
$(x,y)=(r-j,j)$ and put $v=(r-j-1,j)$. The four-spin term becomes
\begin{equation}
 X_{r-1,j}Z_{r,j}X_{r+1,j+1}Z_{r,j+1}
 =X_vZ_{v+e_x}X_{v+e_x+e_y}Z_{v+e_y}.
\end{equation}
A uniform spin rotation $R=\prod_i\exp(-i\pi X_i/4)$ leaves the field
operators unchanged and maps this term to
\begin{equation}
 X_vY_{v+e_x}X_{v+e_x+e_y}Y_{v+e_y}.
\end{equation}
Thus the regular two-dimensional bulk at $N=2$ is a transverse
Wen-plaquette model, whose relation to Ising chains is already
known~\cite{YuKouWen2008}. The finite-volume comparison must retain the
boundary terms and the periodic identifications induced by the lattice
relabelling.

With periodic longitudinal boundaries, $Q_b^{-1}=Q_b$ and the spin
boundary twist reduces to
\begin{equation}
 \theta_a=(-1)^{\sum_{b\sim a}q_b},
 \qquad q_b\in\{0,1\},
\end{equation}
where $b\sim a$ denotes adjacency in the underlying unoriented
transverse graph. This twist is independent of the transverse edge
orientations. Each fixed-charge block remains a free-fermion problem,
with the appropriate fermion-parity projection. Consequently,
longitudinal closure preserves sectorwise free-fermion solvability at
$N=2$, although the unrestricted open-chain occupation rule is replaced
by parity-resolved spectra. For real couplings the Hamiltonian is
diagonalizable, including at spectral degeneracies. Complex couplings
need not have this property.

\begin{figure*}[t]
\centering
\includegraphics[width=0.88\textwidth]{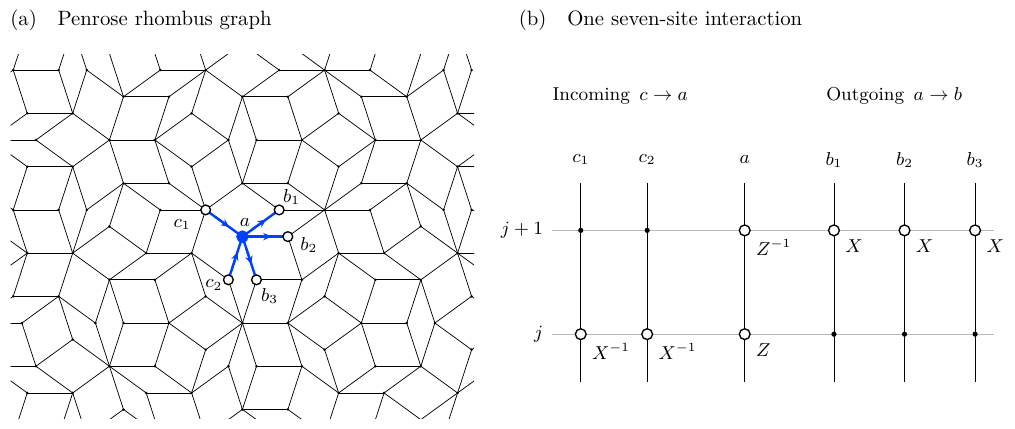}
\caption{
A seven-site interaction in a three-dimensional array of Baxter-Fendley chains perpendicular to a Penrose rhombus tiling. (a) Transverse view showing a $J$-type vertex $a$ and its five neighbours. Arrows indicate the selected orientations \(c_1,c_2\to a\) and \(a\to b_1,b_2,b_3\). (b) Schematic arrangement of the six selected chains at adjacent spatial levels $j$ and $j+1$. The seven open circles carry the operator factors in Eq.~\eqref{eq:penrose-bond}, while the small filled dots mark sites not participating in this term. Vertical lines identify individual chains. The horizontal arrangement in (b) is schematic and does not reproduce the transverse geometry.
(Illustrative diagram: OpenAI Codex (GPT-6) assisted with the TikZ 
code for the diagram, which was reviewed and verified by the authors.)
}
\label{fig:penrose}
\end{figure*}

\section{Baxter meets Penrose}
\label{sec:penrose}

The Penrose rhombus tiling has long-range order but no finite translational unit cell~\cite{Penrose1979}. 
As an application of Section 4, we construct a three-dimensional $Z_N$ clock model with a Penrose tiling as its transverse graph. Take a finite patch $P=(V_P,E_P)$ with open transverse boundaries, as shown in Fig.~\ref{fig:penrose}(a). At each vertex $a$, place a chain of $L$ clock spins perpendicular to the tiling plane. A physical site is labelled by $(a,j)$, where $j$ specifies the third spatial direction. We orient each transverse edge in the direction of increasing $\mathbf u\cdot\mathbf x$, with
$\mathbf u$ perpendicular to no edge.
We take open boundaries along each chain.

To make the quasiperiodic structure explicit in the spectrum, we now specialize the longitudinal couplings according to the local Penrose vertex types. The unitary reduction remains exact, while the open-boundary spectrum is organized by the populations of these types. We use the eight marked vertex types $\alpha\in\{D,Q,K,J,S,S_3,S_4,S_5\}$ of de Bruijn~\cite{deBruijn1981I}.
The same local environments recur throughout the tiling. We assume
that vertices of the same type $\alpha$ have the same couplings
$h_j^{(\alpha)}$ and $J_j^{(\alpha)}$ at a given layer $j$. These
couplings may be complex and may depend on $j$. Each vertex inherits its type from
the full tiling, including at the boundary of $P$.
For a vertex $a$ of type $\alpha$, define
\begin{equation}
\begin{aligned}
 H_\alpha(a)
 &=\sum_{j=1}^{L}h_j^{(\alpha)}X_{a,j}
   +\sum_{j=1}^{L-1}J_j^{(\alpha)}C_{a,j}B_{a,j},\\
 B_{a,j}&=Z_{a,j}Z_{a,j+1}^{-1},\\
 C_{a,j}&=\prod_{b:\,a\to b}X_{b,j+1}
          \prod_{c:\,c\to a}X_{c,j}^{-1}.
\end{aligned}
\label{eq:penrose-vertex}
\end{equation}
Writing $V_\alpha(P)$ for the set of vertices of type $\alpha$ in $P$,
the Hamiltonian of the array is
\begin{equation}
 H_P=\sum_\alpha\sum_{a\in V_\alpha(P)} H_\alpha(a).
\label{eq:penrose-hamiltonian}
\end{equation}
Each edge of $P$ incident on $a$ contributes a factor of $X$ or $X^{-1}$
to $C_{a,j}$. At a vertex of coordination number $z_a$, the bond
interaction $C_{a,j}B_{a,j}$ acts on $z_a+2$ sites.
Table~\ref{tab:penrose-interactions} lists the seven geometrically distinct interior vertex configurations, 
with the marked types $S$ and $S_5$ grouped together.
The Penrose construction therefore has interactions involving at most nine sites.
For $p$ incoming edges, the term contains
$p$ factors of $X^{-1}$ in layer $j$ and $z_a-p$ factors of $X$ in layer
$j+1$, in addition to $B_{a,j}$. At the boundary, $z_a$ counts only the
edges retained in $P$.

\begin{table}[htb]
\centering
\begin{tabular}{@{}lccc@{}}
\hline
Vertex type & $z_a$ & Incoming edges $p$ & Sites in term \\
\hline
$D$        & 3 & 1, 2       & 5 \\
$Q$        & 3 & 1, 2       & 5 \\
$K$        & 4 & 1, 2, 3    & 6 \\
$J$        & 5 & 1, 2, 3, 4 & 7 \\
$S$, $S_5$ & 5 & 2, 3       & 7 \\
$S_4$      & 6 & 2, 3, 4    & 8 \\
$S_3$      & 7 & 3, 4       & 9 \\
\hline
\end{tabular}
\caption{Bond interactions at interior vertices. The possible numbers
$p$ of incoming edges are obtained by counting edge directions in the
open half-plane $\boldsymbol u\cdot(\boldsymbol x_b-\boldsymbol x_a)<0$
as the vertex is rotated relative to $\boldsymbol u$. Types $S$ and $S_5$
have the same arrangement of incident rhombi and differ in their tiling
matching arrows~\cite{BoroczkyRetiWintsche2006}.}
\label{tab:penrose-interactions}
\end{table}

For example, the marked $J$ vertex $a$ in Fig.~\ref{fig:penrose}(a) has
$z_a=5$ and $p=2$. 
Each incoming edge $c_i\to a$ contributes
$X_{c_i,j}^{-1}$, and each outgoing edge $a\to b_i$ contributes
$X_{b_i,j+1}$. Multiplying these five factors by the longitudinal bond
$B_{a,j}=Z_{a,j}Z_{a,j+1}^{-1}$ gives the seven-site operator
\begin{equation}
\begin{aligned}
 C_{a,j}B_{a,j}={}&X_{c_1,j}^{-1}X_{c_2,j}^{-1}
             Z_{a,j}Z_{a,j+1}^{-1}\\
             &{}\times X_{b_1,j+1}X_{b_2,j+1}X_{b_3,j+1}.
\end{aligned}
\label{eq:penrose-bond}
\end{equation}

By Eq.~\eqref{eq:d-interior-bond}, $U^\dagger H_\alpha(a)U$ is the
open Baxter-Fendley Hamiltonian on chain $a$ with couplings
$h_j^{(\alpha)},J_j^{(\alpha)}$. Thus each vertex contributes an
independently chosen one-dimensional energy $e_{\alpha,\nu}$ from
Eq.~\eqref{eq:d-open-spectrum}.

Deflation is the standard Penrose subdivision procedure in which each marked rhombus is replaced by a prescribed arrangement of smaller marked rhombi, according to de Bruijn's rules~\cite{deBruijn1981II}.
To follow successive deflations, fix a finite region and let
$n_\alpha^{(k)}$ count its vertices of type $\alpha$ after $k$
deflations, including vertices on its boundary. In de Bruijn's
construction the old vertices are retained with new types, each thick
rhombus produces one new $J$ vertex, and each distinct red edge of the marked tiling
produces one new $D$ vertex~\cite{deBruijn1981II}. Applying these rules
and counting the vertices in the chosen region gives the multiplicities
at each step. 
The red arrows are the matching marks of the tiling. The arrows in Fig.~\ref{fig:penrose}(a) specify the orientations used in the Hamiltonian.
For example, consider the closed patch formed by five thick rhombi meeting at an $S$ vertex. Its eleven original vertices comprise one $S$, five $D$ and five $J$ vertices. Under one deflation, these become one $S_5$, five $Q$, and five $K$ vertices. The ten outer edges of this five-rhombus patch are red-arrow edges in de Bruijn's marked-tiling convention. Under deflation, each produces one new $D$ vertex, while each of the five thick rhombi produces one new $J$ vertex. Including the boundary, the deflated patch therefore contains 26 vertices.
Thus the vertex counts entering Eq.~\eqref{eq:penrose-deflation-energy} are one $S_5$, five $Q$, five $K$, ten $D$ and five $J$ vertices.

For the vertex populations $n_\alpha^{(k)}$ at deflation stage $k$, the corresponding energies are
\begin{equation}
 E^{(k)}=\sum_\alpha\sum_{r=1}^{n_\alpha^{(k)}}
                  e_{\alpha,\nu_{\alpha,r}},
\label{eq:penrose-deflation-energy}
\end{equation}
where each $\nu_{\alpha,r}$ is chosen independently. If every vertex
of type $\alpha$ is assigned the same level $e_\alpha$, this gives
$E^{(k)}=\sum_\alpha n_\alpha^{(k)}e_\alpha$.

\section{Conclusion}

For transverse graphs with uniformly bounded edge lengths and vertex degrees, the construction gives finite-range multispin Hamiltonians. Rectangular lattices and the Penrose array are particular examples. Each longitudinal bond interaction acts on $z_a+2$ sites, where $z_a$ is the transverse coordination number. With open longitudinal boundaries, the nonlocal unitary $U$ reduces the Hamiltonian to independent Baxter-Fendley chains. Arrays with the same chain lengths and longitudinal couplings therefore have identical spectra, counted with algebraic multiplicity, irrespective of their transverse connectivity.

With periodic longitudinal boundaries, the twist on each closing bond
depends on the difference between the sums of the outgoing and incoming
neighbours' charge labels. In each joint charge sector, every array energy is a sum of the corresponding twisted periodic-chain energies. All sectors together give the full spectrum.
A natural direction for future work is to investigate original-spin correlations and, 
for periodic longitudinal boundaries, graph-dependent ground-state degeneracies, 
particularly in the Ising limit on the Penrose array.

\section*{Acknowledgments}

M. T. B. gratefully acknowledges the hospitality and support of the Asia Pacific Center for Theoretical Physics (APCTP), Pohang, during the completion of this work.

\section*{Declaration of generative AI and AI-assisted technologies in the manuscript preparation process}

During the preparation of this work, the authors used OpenAI Codex (GPT-6) to assist with generating TikZ code for illustrative diagrams, and ChatGPT to assist with literature searches, the organisation and review of mathematical notes, and suggestions concerning manuscript wording and presentation. All AI-assisted material was reviewed and edited by the authors. The mathematical arguments, references and figures were independently checked by the authors, who take full responsibility for the content of the published article.

\end{document}